\documentclass{ws-ijmpa}
\usepackage[super,compress]{cite}
\usepackage{graphicx}
\usepackage{fancybox, ascmac}
\begin{document}
\markboth{H. Sakuma et al.}{Conserved relativistic Ertel's current}

%
\catchline{}{}{}{}{}
%


\title{Conserved relativistic Ertel's current generating the vortical and thermodynamic aspects of spacetime}


\author{Hirofumi Sakuma\footnote{
Research Origin for Dressed Photon, 3-13-19 Moriya-cho, Kanagawa-ku,
Yokohama-shi, Kanagawa 221-0022, Japan},$\;$$^{{\dag}}$ Izumi Ojima,$^{*}$$^{{\S}}$,
Hayato Saigo$^{\ddag}$$^{\sharp}$ and Kazuya Okamura$^{*,\#}$$^{\P}$}


\address{$^{*}$Research Origin for Dressed Photon, \\
Yokohama, Kanagawa 221-0022, Japan\\
$^{\ddag}$Nagahama Institute of Bio-Science and Technology,\\
1266 Tamura, Nagahama, Shiga 526-0829, Japan \\
$^{\#}$ Graduate School of Informatics, Nagoya University \\
Chikusa-ku, Nagoya 464-8601, Japan\\
\textnormal{$^{\dag}$}sakuma@rodrep.or.jp, \textnormal{$^{\S}$}ojima@gaia.eonet.ne.jp \\
\textnormal{$^{\sharp}$}h\_saigoh@nagahama-i-bio.ac.jp, \textnormal{$^{\P}$}k.okamura.renormalizable@gmail.com
}



\maketitle


\centerline{{\fontsize{8pt}{3pt}\selectfont \it Submitted to Int. J. Mod. Phys. A}}
\vspace{5mm}
\begin{abstract}

Motivated by Aoki et al.\rq s recent research on conserved charges and entropy current, we investigated the conservation of relativistic Ertel\rq s current, which has received little attention outside the field of geophysical fluid dynamics. Ertel\rq s charge is an important indicator of the correlation between vortex vectors and entropy gradient fields in Earth\rq s meridional heat transport. We first show that in the generalized Hamiltonian structure of baroclinic fluids, the duality between the total energy and the Casimir as a function of Ertel\rq s charge plays an important role in the nonrelativistic case. Then, by extending the result to relativistic cases, we show that this finding has far-reaching implications not only for spacetime issues in cosmology but also for the foundation of quantum field theory. An especially important finding is that, as an unreported dual form of the Einstein field equation, we identify a special equation satisfied not only by the vortex tensor field generated by the conserved charge but also by the Weyl tensor in interpreting the physical nature of the metric tensor $g^{\mu\nu}$, which appears in the cosmological term $\Lambda g^{\mu\nu}$.

\end{abstract}

\ccode{04.20.Cv; 04.40.Nr; 05.90.+m; 47.10.ab}


\section{Introduction}

Two key issues are addressed in this article: first, we shed new light on the peculiarity of the gravitational field, which has, in addition to its well-documented geometrical nature, an unexpected thermodynamic nature \cite{jb1,swh1,wgu, ver}; second, we clarify the main cause of the inherent defect in the present form of quantum field theory (QFT), exemplified by Haag\rq s no-go theorem\cite{sw} in axiomatic QFT. As we show below, these two seemingly disparate concepts are inextricably linked by the important dynamical role of the spinor field (or the vortical field in the case of classical mechanics), which connects a given physical system to the surrounding spacetime as its dynamical environment.

Many issues in nonrelativistic physics require the use of an abstract space to characterize the time-dependent dynamic behaviors of a given physical system, such as the Hilbert space $\mathfrak{H}$ in quantum mechanics and phase space in classical Hamiltonian ($H$) systems. In general, these spaces are not related to the physics of a given system. The fact that this condition changes drastically in relativistic field theory is well known and has become common knowledge in physics, while the well-known \textit{wave-particle duality} in quantum mechanics remains a mystery. However, once it is accepted that spacetime is a physical entity in which \lq\lq spacelike\rq\rq corresponds to wavelike entities, such as the spacelike momentum field required for quantum field interactions (Greenberg-Robinson (GR) theorem\cite{jr, da}), while \lq\lq timelike\rq\rq corresponds to localized particle-like entities, then the wave-particle duality of quantum entities appears to be a natural consequence of the fact that quantum entities always coexist with embedded spacetime; in other words, quantum entities are dynamically integrated with spacetime as their environment.

Dirac's finding has played an important role in considering the physicality of spacetime, revealing that quantum spin is a feature of relativistic spacetime. Since then, the idea that spacetime is composed of specific types of spin networks has been investigated, most notably by Penrose\cite{rp1} and the many researchers that developed loop quantum gravity (LQG) theory. Therefore, an essential question associated with the spacetime conundrum is how the thermodynamic properties of spacetime fit with this spin network concept. The peculiarity of the spacetime problem in relativistic scenarios lies in the fact that spacetime not only has a mathematical meaning, which can be used to represent a given physical system, but is also a physical field that must be represented itself.
To explain this reciprocal nature and the interdependent duality (ID) that exists between a given physical system and its associated spacetime, we must identify the emergent processes of this \lq\lq physical\rq\rq spacetime.

As an informative example of the ID mentioned above, we briefly refer to micro-macro duality (MMD) theory \cite{ojima1}, which, as briefly explained in a recent review paper \cite{sak1} on nanophotonics, {\it can be regarded as a modern version of \lq\lq quantum-classical correspondence\rq\rq; this theory was rigorously derived based on the pertinent generalization of the superselection rule, with the sectors originally formulated by Doplicher-Haag-Roberts \cite{dhr1, dr1}}. Any efforts to explore unseen microscopic quantum worlds require experiments that investigate the dynamical interactions between micro-quantum and macro-classical worlds. This simple fact clearly shows that our descriptions of unseen microscopic worlds depend heavily on the \lq\lq vocabulary\rq\rq we use in classical physics. Therefore, in the metaphorical sense, classical physics operates not as a static \lq\lq space\rq\rq onto which the microscopic world is projected, but rather as an interacting dynamic \lq\lq spacetime\rq\rq which we use to describe the targeted microscopic world. 

This viewpoint is supported as a key element of MMD theory. Among the many noteworthy accomplishments of MMD theory, the following two theoretical results are particularly relevant to the discussion in this paper. First, due to its infinite degrees of freedom, QFT inevitably involves a combination of quantum and classical fields, with the latter emerging due to the existence of disjoint (refined notion of unitary non-equivalent) generalized sectors with non-trivial factor representations acting as order parameters of the emerging classical field. The irrelevance of Schr\"{o}dinger\rq s cat thought experiment can be easily clarified\cite{sak1} by the basic results of MMD theory. Second, with the exception of the different algebraic structures of certain physical quantities in their respective fields, i.e., anti-commutativity vs. commutativity, we can consider both fields in a unified framework.

The main discussions begin in Section 2 with classical physics; however, we show that even in classical fluid dynamics systems, there exists an intriguing ID structure that eventually clarifies the discussed spacetime peculiarity. In Sections 2 and 3, by reviewing the basic structures of barotropic and baroclinic fluids, as well as their generalized Hamiltonian ($H$) structures, we identify a key conserved dynamical quantity that generates the abovementioned ID structure. The reason we focus on this conserved quantity is because we found that recent works by Aoki et al.\cite{aoki1, aoki2}, particularly on the conservation of entropy in general relativity (GR), are quite informative in the present research. Then, in Section 4, we show how the main aims of our paper, as stated in the first paragraph of this introductory section, are achieved based on the results of the previous section. In the final section, we present our conclusions and some novel perspectives on QFT and cosmology.

\section{Dynamics of barotropic and baroclinic fluids and $H$ structures}
We begin with the well-known equation of motion of a perfect fluid in nonrelativistic fluid mechanics:
\begin{equation}
D_{t}v_{\mu}:=\partial_{t}v_{\mu}+v^{\nu}\partial_{\nu}v_{\mu}=-\frac{1}{\rho}\partial_{\mu}p,  \label{eqn:1}
\end{equation} 
where the notations are conventional. A fluid is classified as barotropic or baroclinic based on its form of $\partial_{\mu}p/\rho$; that is, if $p=p(\rho)$ or $\rho=\rho_{0}=const.$, the fluid is barotropic; otherwise, the fluid is baroclinic. In other words, the barotropic or baroclinic nature of the fluid is characterized by whether $\partial_{\mu}p/\rho$ is a conservative force field, which has a decisive influence on whether the associated vorticity field is conservative. Thus, we refer to the \textit{Langange\rq s vortex theorem for barotropic flows}, which shows that vortices are free of generation and extinction.

Since we are primarily interested in vorticity and entropy fields, it is useful to rewrite the baroclinic form of Eq. (\ref{eqn:1}) in terms of the vorticity $\zeta_{\mu\nu}$ and specific (i.e., per unit mass) entropy $s$ fields by using the first law of thermodynamics (Eq. (\ref{eqn:3})) and the vector identity given in Eq. (\ref{eqn:5}):
\begin{eqnarray}
-dp/\rho = Tds - dw,\;\textnormal{with}\;\; \partial_{t}s + v^{\nu}\partial_{\nu}s =0,   \label{eqn:3} \\
v^{\nu}\partial_{\nu}v_{\mu}=v^{\nu}(\partial_{\nu}v_{\mu}-\partial_{\mu}v_{\nu})+\partial_{\mu}(v^{\nu}v_{\nu}/2),  \label{eqn:5}
\end{eqnarray}
where $T$ and $w$ are the absolute temperature and specific enthalpy, respectively. From Eqs. (\ref{eqn:3}) and (\ref{eqn:5}), Eq. (\ref{eqn:1}) becomes
\begin{equation}
\partial_{t}v_{\mu}+\partial_{\mu}(w+v^{\nu}v_{\nu}/2)-\zeta_{\mu\nu}v^{\nu}=T\partial_{\mu}s.  \label{eqn:7}
\end{equation}

The most well-known example of a baroclinic fluid is the atmosphere, for which the ideal gas law can be applied with a high degree of accuracy. The atmosphere is particularly important in our discussion because it provides a useful fluid dynamic system with a nonuniform entropy distribution in both the vertical and meridional directions; furthermore, energetic vortical fields known as baroclinic eddies play important dynamical roles in heat transport along the meridional direction. In the dynamics of this heat transport, there is a strong correlation between the vorticity $\vec{\zeta}$ and entropy gradient $\vec{\nabla}s$ fields, which can be described by Ertel\rq s potential vorticity $Q$ \cite{ert}, which is defined as
\begin{equation}
Q:= \frac{1}{\rho}(\vec{\zeta}\cdot\vec{\nabla}s),\;\;\partial_{t}Q+v^{\nu}\partial_{\nu}Q=0. \label{eqn:9}
\end{equation}
The above is the most important conserved quantity in the field of geophysical fluid dynamics.

It should be noted that the importance of baroclinicity in the atmosphere varies with scale. In general, in typical laboratory experiments using air, such as wind tunnel studies, air flows behave as barotropic flows since the entropy gradient is negligible; however, for air flows with horizontal scales greater than several hundred or a few thousand kilometers, baroclinicity becomes a nonnegligible dynamical factor. In the introduction, we discussed MMD theory, which indicates that the actual world in which we live consists of an ID structure that bridges quantum and classical physics, rather than \textit{ the prevailing view that the laws of classical physics are not fundamental but are approximated from \lq\lq genuine fundamental\rq\rq quantum mechanics}. We believe that the scale dependency of baroclinicity, represented by $Q$ in the comparison of barotropic and baroclinic flows, is analogous to that of the Planck constant $\bar{h}$ in the comparison of quantum and classical physics in MMD theory. To show that this resemblance is not superficial but rather has essential implications for the main issue in this paper, we investigated the generalized $H$ structure of baroclinic flows.

The generalized $H$ structure was derived from the so-called non-canonical form of the $H$ formulation, with the seminal work initiated by Arnol\rq d \cite{arn} and further developed by a group of applied mathematicians and physicists \cite{mg, ddh1, mm}. For a baroclinic perfect fluid dynamics system with Eulerian representation, the generalized $H$, denoted by $H_{G}$, has the form\cite{kur}:
\begin{equation}
H_{G}:= E+C_{F},\;\;\;E:= \int\langle \rho v^{\nu}v_{\nu}/2 + \rho e(\rho, s) \rangle dV,\;\;\;C_{F}:= \int\langle \rho F(s, Q) \rangle dV,   \label{eqn:11}
\end{equation}
where $E$ and $C_{F}$ are the total energy with $e(\rho, s)$ being the internal energy density, a Casimir constructed by an arbitrary function $F$ of $s$ and $Q$. Since the given fluid dynamical system can be described by five independent variables, namely, $v^{\mu}, (1\leq \mu\leq 3)$ and two thermodynamical variables, we choose $\rho$ and $s$ as the thermodynamic variables because $Q$ is expressed in terms of these two variables. First, when we compare the conservative quantity $C_{F}$ with the total energy $E$, we can observe that it is not merely an additional constant of motion. This occurs because both $E$ and $C_{F}$ are \lq\lq complete\rq\rq in the sense that they include all five variables. In other words, they are equal pairs of \lq\lq complete\rq\rq constants of motion.

The significant advantage of $H_{G}$ over $E$ becomes clear when we consider the stability of a given steady state of the fluid because any given steady state of the baroclinic flow can be represented by the condition that the first variation in $H_{G}$ vanishes; that is, 
\begin{equation}
\delta H_{G} = \delta E + \delta C_{F} = 0, \label{eqn:13}
\end{equation}
which can be rewritten as the combination of the steady-state version of Eq. (\ref{eqn:7}) and
\begin{equation}
F-Q\frac{\partial F}{\partial Q}+B(s, Q) =0.    \label{eqn:15}
\end{equation}
Eq. (\ref{eqn:15}) shows that an arbitrary function $F$ can be determined by Bernoulli\rq s function $B(s, Q)$, which characterizes the given steady state. Then, we can demonstrate the formal stability of the state \cite{sak2} (the stability of a given steady state for infinitesimally small amplitude perturbations with arbitrary forms) if the second variation in $H_{G}$ is sign definite. In our discussion of the ID structure, a particularly important aspect of Eq. (\ref{eqn:13}) is that the balance between $\delta E$ and $\delta C_{F}$ can be regarded as a unique \lq\lq interaction\rq\rq between two dynamics with and without explicit forms of $Q$. It is clear that the phase-space trajectory of a linearly unstable mode, such as those represented by one of the separatrices of $H_{G}$, i.e., $\delta^{2}H_{G}=0$, cannot be described without $C_{F}$. Thus, we believe that perturbation methods with only the Hamiltonian as the total energy are not generally adaptable since, as our present discussion shows, these methods cannot cover the dynamic behaviors of certain unstable modes. We also believe that the root cause of Haag\rq s no-go theorem in axiomatic QFT, as discussed in the introduction, can be attributed to such an inability inherent in the conventional Hamiltonian approaches. To further show the decisive role of $Q$ in ID structures, we next examine the relativistic expression of $Q$.

\section{Converted form of the relativistic equation of motion}
First, we fix the sign convention as $(+, -, -, -)$ and introduce a nondimensional four-velocity vector $u^{\mu}$ that satisfies the normalization condition $u^{\nu}u_{\nu}=1$. In this section and Sections 3 and 4, for simplicity, unless otherwise stated, we develop our arguments within the framework of special relativity. The following basic arguments on the relativistic equations from Eqs. ((\ref{eqn:17})-(\ref{eqn:27})) are given by Landau \& Lifshitz \cite{ll} (pp. 506-508). Regarding the first law of thermodynamics, we have\begin{equation}
-\frac{dp}{n} = Td\left(\frac{\sigma}{n}\right) - d\left(\frac{w}{n}\right),  \label{eqn:17} 
\end{equation}
where $n$ denotes the \lq\lq particle number\rq\rq corresponding to the density $\rho$ in the nonrelativistic case, and $\sigma/n$ and $w/n$ are the specific entropy and enthalpy, respectively, as in Eq. (\ref{eqn:3}).
The energy-momentum tensor for a perfect fluid has the following form:\begin{equation}
T^{\mu\nu} = w u^{\mu}u^{\nu} - p g^{\mu\nu}.  \label{eqn:19}
\end{equation}
The tensor divergence of Eq. (\ref{eqn:19}) gives the following equations of motions:
\begin{equation}
\partial_{\nu}T_{\mu}^{\;\;\nu} = u_{\mu}\partial_{\nu}(w u^{\nu}) +wu^{\nu}\partial_{\nu}u_{\mu}- \partial_{\mu}p =0.  \label{eqn:21}
\end{equation}
The projection of Eq. (\ref{eqn:21}) in the direction $u^{\mu}$ can be calculated by $u^{\mu}\partial_{\nu}T_{\mu}^{\;\;\nu}=0$; with Eq. (\ref{eqn:17}), this becomes
\begin{equation}
u^{\nu}\partial_{\nu}(\sigma/n) = 0,   \label{eqn:23}
\end{equation} 
which corresponds to the second equation in Eq. (\ref{eqn:3}). Next, we calculate the component of $\partial_{\nu}T_{\mu}^{\;\;\nu}=0$ perpendicular to $u^{\mu}$ as \begin{equation}
\partial_{\nu}T_{\mu}^{\;\;\nu} - u_{\mu}u^{\nu}\partial_{\sigma}T_{\nu}^{\;\;\sigma}=0,  \label{eqn:25}
\end{equation}
which yields
\begin{equation}
w u^{\nu}\partial_{\nu}u_{\mu}-\partial_{\mu}p+u_{\mu}u^{\nu}\partial_{\nu}p =0.   \label{eqn:27}
\end{equation}
This equation corresponds to the nonrelativistic form of $\rho D_{t}v_{\mu}=0$ in Eq. (\ref{eqn:1}).

As we have noted with Eq. (\ref{eqn:7}) in Section 2, in terms of the vorticity and the entropy field, the form of Eq. (\ref{eqn:7}) is preferable to the form of Eq. (\ref{eqn:1}).
The derivation of the relativistic form of Eq. (\ref{eqn:7}) is straightforward and has the following form: 
\begin{equation}
\omega_{\mu\nu}u^{\nu}=T\partial_{\mu}(\sigma/n),\;\;\omega_{\mu\nu}:=\partial_{\mu}[(w/n)u_{\nu}]
-\partial_{\nu}[(w/n)u_{\mu}].   \label{eqn:29}
\end{equation}

Appendix A includes the details of this derivation since, to the best of our knowledge, there are no references that describe this derivation. Eq. (\ref{eqn:23}) is included in Eq. (\ref{eqn:29}), as shown by $0=u^{\mu}\omega_{\mu\nu}u^{\nu}=Tu^{\mu}\partial_{\mu}(\sigma/n)$. We can also observe that Eq. (\ref{eqn:29}) remains valid for curved spacetime if we replace $\partial_{\mu}$ with a covariant derivative, denoted as $\nabla_{\mu}$. In particular, we have that $\omega_{\mu\nu}=\partial_{\mu}v_{\nu}-\partial_{\nu}v_{\mu}=\nabla_{\mu}v_{\nu}-\nabla_{\nu}v_{\mu}$.

\section{Ertel's charge as an entropic vortex field}
To appreciate the importance of Eq. (\ref{eqn:29}), we use an explicit (writing down all the elements) matrix representation of Eq. (\ref{eqn:31}) below to derive the conserved Ertel's current.
\begin{eqnarray}
\qquad \omega_{\mu \nu}=\left( 
\begin{array}{cccc}
0 & \omega_{01} & \omega_{02} & \omega_{03} \\ 
-\omega_{01} & 0 & \omega_{12} & -\omega_{31} \\ 
-\omega_{02} & -\omega_{12} & 0 & \omega_{23} \\ 
-\omega_{03} & \omega_{31} & -\omega_{23} & 0  \label{eqn:31} 
\end{array}
\right). 
\end{eqnarray}
First, after defining the pseudoscalar $\Omega$ in Eq. (\ref{eqn:33}), we introduce $\hat{\omega}^{\mu\nu}$, which is the Hodge dual of $\omega_{\mu \nu}$, i.e.,
\begin{equation}
\Omega:=\omega_{01}\omega_{23}+\omega_{02}\omega_{31}+\omega_{03}\omega_{12}, \label{eqn:33} 
\end{equation}
\begin{eqnarray}
\qquad \hat{\omega}^{\mu \nu}=\left( 
\begin{array}{cccc}
0 & -\omega_{23} & -\omega_{31} & -\omega_{12} \\ 
\omega_{23} & 0 & -\omega_{03} & \omega_{02} \\ 
\omega_{31} & \omega_{03} & 0 & -\omega_{01} \\ 
\omega_{12} & -\omega_{02} & \omega_{01} & 0   \label{eqn:35} 
\end{array}
\right). 
\end{eqnarray}
From Eqs. (\ref{eqn:31}) and (\ref{eqn:35}), we obtain\begin{equation}
\hat{\omega}^{\mu\kappa}\omega_{\kappa\nu}=\Omega g^{\mu}_{\;\;\nu},\;\;\;
\hat{\omega}^{\mu\nu}\omega_{\nu\mu}=4\Omega.    \label{eqn:37}
\end{equation}
According to Eq. (\ref{eqn:29}), we have
\begin{equation}
(\hat{\omega}^{\mu\kappa}\omega_{\kappa\nu})u^{\nu}=\Omega g^{\mu}_{\;\;\nu}u^{\nu}=T\hat{\omega}^{\mu\kappa}\partial_{\kappa}(\sigma/n);   \label{eqn:39} 
\end{equation}
thus, we can obtain 
\begin{equation}
\Omega_{T} u^{\mu}=\hat{\omega}^{\mu\kappa}\partial_{\kappa}(\sigma/n),\;\;\textnormal{where}\;\;\Omega_{T}:= \Omega/T.  
\label{eqn:41}
\end{equation}
By substituting Eq. (\ref{eqn:35}) into Eq. (\ref{eqn:41}) and with a series of manipulations based on the skew symmetry of $\omega_{\mu\nu}$, we can finally derive that 
\begin{eqnarray}
\Omega_{T}\left( 
\begin{array}{c}
u^{0} \\
u^{1} \\
u^{2} \\
u^{3}
\end{array} 
\right) =\left( 
\begin{array}{c}
-\partial_{1}[\omega_{23}(\sigma/n)]-\partial_{2}[\omega_{31}(\sigma/n)]-\partial_{3}[\omega_{12}(\sigma/n)]  \\ 
\partial_{0}[\omega_{23}(\sigma/n)]-\partial_{2}[\omega_{03}(\sigma/n)]+\partial_{3}[\omega_{02}(\sigma/n)] \\ 
\partial_{0}[\omega_{31}(\sigma/n)]+\partial_{1}[\omega_{03}(\sigma/n)]-\partial_{3}[\omega_{01}(\sigma/n)] \\ 
\partial_{0}[\omega_{12}(\sigma/n)]-\partial_{1}[\omega_{02}(\sigma/n)]-\partial_{2}[\omega_{01}(\sigma/n)]  \label{eqn:43} 
\end{array}
\right). 
\end{eqnarray}
Based on this expression, we can observe that
\begin{equation}
\partial_{\nu}(\Omega_{T}u^{\nu})=0.   \label{eqn:45}
\end{equation}

However, according to Eq. (\ref{eqn:29}), we have
\begin{equation}
-\Omega u^{0} = T[\omega_{23}\partial_{1}(\sigma/n)+\omega_{31}\partial_{2}(\sigma/n)+\omega_{12}\partial_{3}(\sigma/n)]; \label{eqn:47}
\end{equation}
thus, we obtain 
\begin{equation}
\Omega_{T}=-[\omega_{23}\partial_{1}(\sigma/n)+\omega_{31}\partial_{2}(\sigma/n)+\omega_{12}\partial_{3}(\sigma/n)]/u^{0},   \label{eqn:49}
\end{equation}
which is the relativistic expression of Ertel\rq s potential vorticity $Q$ given in Eq. (\ref{eqn:9}). According to theoretical physics convention, we refer to $\Omega_{T}$ as Ertel\rq s charge. The conservation property of Eq. (\ref{eqn:45}) can be extended to curved spacetime by substituting $\partial_{\mu}$ in Eq. (\ref{eqn:43}) with the covariant derivative $\nabla_{\mu}$ and using the tensor identity\begin{equation}
(\nabla_{\mu}\nabla_{\nu}-\nabla_{\nu}\nabla_{\mu})\omega_{\kappa\lambda} = -R^{\sigma}_{\;\kappa\mu\nu}\omega_{\sigma\lambda}-R^{\sigma}_{\;\lambda\mu\nu}\omega_{\kappa\sigma}  \label{eqn:51}
\end{equation}
to calculate the vector divergence on the right side of Eq. (\ref{eqn:43}), where $R^{\alpha}_{\;\;\beta\gamma\delta}$ denotes the Riemann curvature tensor. According to the second equation in Eq. (\ref{eqn:29}), the physical dimension of $\Omega$ in Eq. (\ref{eqn:33}), denoted by $dim[\Omega]$, becomes $dim[\Omega]=l^{-2}dim[(w/n)^{2}]$, where $l$ denotes the length scale. Since $n$ and $w$ are the particle number and the energy per unit volume, respectively, if we use a natural unit system, then $dim[n]=l^{-3}$ and $dim[w^{2}]=l^{-8}$. Thus, $dim[\Omega]=l^{-1}/l^{3}$, indicating that $dim[\Omega]=dim[T^{\mu\nu}]$ in Eq. (\ref{eqn:19}) and hence, from Eq. (\ref{eqn:17}), \textit{the physical dimension of $\Omega_{T}$ is the entropy per unit volume.}

A particularly intriguing property of $\omega_{\mu\nu}$ is that \lq\lq Dirac\rq s $\gamma$ matrix\rq\rq $\;\hat{\gamma}^{\mu\nu}$ can be constructed from Eq. (\ref{eqn:37}). In fact, if we define $\hat{\gamma}^{\mu}_{\;\nu}$ such that $\hat{\gamma}^{\mu}_{\;\nu}:=\hat{\omega}^{\mu\sigma}\omega_{\sigma\nu}$, then, according to Eq. (\ref{eqn:37}), $\hat{\gamma}^{\mu}_{\;\nu}=\Omega g^{\mu}_{\;\nu}$. By raising the suffix $\nu$, we have that $\hat{\gamma}^{\mu\nu}=\Omega g^{\mu\nu}$; thus, we find that 
\begin{equation}
\frac{1}{\Omega}(\hat{\gamma}^{\mu\nu}+\hat{\gamma}^{\nu\mu})=2g^{\mu\nu},  \label{eqn:53}
\end{equation} 
which is the well-known anti-commutation relation. To further examine the implications of Eq. (\ref{eqn:53}), we investigated the relation between $g^{\mu\nu}$ and $\hat{\omega}^{\mu\sigma}\omega_{\sigma}^{\;\nu}/\Omega$. According to Eq. (\ref{eqn:37}), $g^{\mu\nu}$ can be rewritten as follows:
\begin{equation}
g^{\mu\nu}=\frac{\hat{\omega}^{\mu\sigma}\omega_{\sigma}^{\;\;\nu}}{\Omega}=\frac{\hat{\omega}^{\mu\sigma}\omega_{\sigma}^{\;\;\nu}(\hat{\omega}^{\kappa\lambda}\omega_{\lambda\kappa})}{\Omega(\hat{\omega}^{\kappa\lambda}\omega_{\lambda\kappa})}=\frac{\hat{\omega}^{\mu\sigma}\omega_{\sigma}^{\;\;\nu}(\hat{\omega}^{\kappa\lambda}\omega_{\lambda\kappa})}{(\hat{\omega}^{\kappa\lambda}\omega_{\lambda\kappa})^{2}/4}.   \label{eqn:55}
\end{equation}
Recall that, in general, , $g^{\mu\nu}$ is not a physical quality but rather a purely mathematical quantity. However, there exists an exceptional case in which $g^{\mu\nu}$ becomes physical, as shown by 
Eq. (\ref{eqn:57}), which was derived by lengthy straightforward calculations \cite{sak3, sak5} on the Weyl conformal tensor $W_{\alpha\beta\gamma\delta}$.
\begin{equation}
W^{\mu\alpha\beta\gamma}W^{\nu}_{\;\;\alpha\beta\gamma}-\frac{1}{4}W^{2}g^{\mu\nu}=0,\;\;
W^{2}:=W^{\alpha\beta\gamma\delta}W_{\alpha\beta\gamma\delta}.   \label{eqn:57}
\end{equation}
Eq. (\ref{eqn:57}) shows that, for non-vanishing $W^{2}$, the cosmological term $\Lambda g^{\mu\nu}$ can be interpreted not as vacuum but as (conformal) gravitational energy. By comparing Eq. (\ref{eqn:55}) with Eq. (\ref{eqn:57}), we find that 
\begin{eqnarray}
g^{\mu\nu}=\frac{W^{\mu\alpha\beta\gamma}W^{\nu}_{\;\;\alpha\beta\gamma}}{W^{2}/4}
=\frac{\hat{\omega}^{\mu\sigma}\hat{\omega}^{\kappa\lambda}\omega^{\nu}_{\;\;\sigma}\omega_{\kappa\lambda}}{(\hat{\omega}^{\kappa\lambda}\omega_{\lambda\kappa})^{2}/4}
=\frac{\hat{\omega}^{\mu\sigma}\hat{\omega}^{\kappa\lambda}\omega^{\nu}_{\;\;\sigma}\omega_{\kappa\lambda}}{(4\Omega)^{2}/4},    \label{eqn:59} 
\end{eqnarray}
which clearly shows that $(4\Omega)^{2}$ correlates directly with $W^{2}$.

\section{Brief conclusions and novel perspectives on QFT and cosmology}
\subsection{Conclusions on $\Omega_{T}$ dynamics and its implication for QFT}
By investigating the relativistic form of Ertel\rq s charge $\Omega_{T}$, which has been largely ignored except in the field of geophysical fluid dynamics, we found that $\Omega_{T}u^{\mu}$ is a conserved \lq\lq entropy current\rq\rq. The importance of this finding is that while the physical dimension of $\Omega_{T}$ is the entropy per unit volume, this quantity is not identical to the thermodynamic entropy density; instead, it is related to both the vortical modes of a given energy-momentum field $T^{\mu\nu}$ and the associated spacetime $g^{\mu\nu}$ (defined within the framework of conformal gravity (\ref{eqn:57})). As a result, Ertel\rq s charge $\Omega_{T}$ plays an important role in the ID structure discussed in the introduction. 

We think that the category-theoretic perspective plays a crucial role in exploring the nature of Ertel\rq s charge in terms of QFT. In thermodynamics, the concept of entropy is understood through the order structure between thermodynamic states and the state transitions between them. Categories are a generalization of both the order and the group-theoretic structure, and they are helpful in capturing the essence of the spatio-temporal structure. In this respect, Saigo \cite{saig} proposed the idea of considering the category algebra, which is a noncommutative convolution algebra defined on a spatiotemporal category; furthermore, he considered the states as linear functions on it, as quantum fields and their states. We believe that by reformulating the discussion of Ertel\rq s charge from this perspective, we can naturally \lq\lq quantize\rq\rq the contents of this paper. In the context of the quantized theory, the (possibly continuous) sector structure arises, which describes the macroscopic nature of the theory. As we have referred to in the brief explanation of MMD theory given in the introductory section, so-called {\it order parameters} distinguish different sectors which have been shown to be treatable in recent quantum measurement theory\cite{ok21}. We believe that, as a future challenge, it is important to reexamine the role of Ertel\rq s charge from the viewpoint of sector theory.

\subsection{Novel perspective on cosmology}
\subsubsection{On a dark matter model}
In Section 4, we showed that $dim[\Omega]=dim[T^{\mu\nu}]$ in Eq. (\ref{eqn:19}), and we found that the nonzero value of $\Omega^{2}$ corresponds to the nonzero value of $W^{2}$, which suggests that the nonzero $\Omega$ is associated with a special energy field associated with nonzero $W^{2}$. In general, in the Einstein field equation (\ref{eqn:61})\begin{equation}
R^{\mu\nu}-\frac{1}{2}Rg^{\mu\nu}+\Lambda g^{\mu\nu}=-\frac{8\pi G}{c^{4}}T^{\mu\nu},  \label{eqn:61}
\end{equation}
energy-momentum fields are associated directly with Ricci curvature terms; thus, a peculiar energy field such as $\Omega$ would be related to the cosmological term $\Lambda g^{\mu\nu}$ we discussed with Eq. (\ref{eqn:57}) to some (or a large) extent.
As the first step toward understanding the physical meaning of a conserved \lq\lq entropy\rq\rq density $\Omega_{T}(=\Omega/T)$, we introduce a constant reference temperature $T_{R}$, the magnitude of which is immaterial at this point in our discussion but which will become important in our newly proposed hypothesis known as simultaneous conformal symmetry breaking (SCSB) of electromagnetic and gravitational fields. With $T_{R}$, we can introduce a nondimensional parameter such as the particle number $\tilde{n}:=T_{R}/T$, which is inversely proportional to the temperature $T$. With $\tilde{n}$, we can rewrite Eq. (\ref{eqn:45}) as
\begin{equation}
\nabla_{\nu}(\tilde{n}\Omega u^{\nu}) = 0.   \label{eqn:63}
\end{equation}
Therefore, if we redefine $\tilde{\Omega}$ as $\tilde{\Omega}:=\tilde{n}\Omega$ and introduce $\tilde{T}^{\mu\nu}$ as $\tilde{T}^{\mu\nu}:=\tilde{\Omega}u^{\mu}u^{\nu}$, we can obtain
\begin{equation}
\nabla_{\nu}\tilde{T}^{\mu\nu} = 0,    \label{eqn:65}   
\end{equation}
since $u_{\mu}$ satisfies the geodesic condition $u^{\nu}\nabla_{\nu}u_{\mu}=0$ at the galactic scale. Note that the nonzero $\tilde{\Omega}$ does not correspond to the nonzero Ricci scalar curvature $R$, but instead corresponds to the nonzero Weyl curvature $W^{2}$; thus, the nonzero current $\tilde{\Omega}u^{\mu}$ can exist even in \lq\lq nearly vacuum\rq\rq regions where $R^{\mu\nu} \approx 0$. Furthermore, since the magnitude of $\tilde{\Omega}$ is inversely proportional to $T$, $\tilde{\Omega}$ would become extremely small soon after the beginning phase of the big bang and then increases, which suggests that the current $\tilde{\Omega}u^{\mu}$ in a region where $R^{\mu\nu}\approx 0$ is a promising candidate for the (cold) dark matter model.

\subsubsection{Brief review of previous cosmological studies by Sakuma et al.}
In their recent study on the off-shell properties of quantum fields motivated by enigmatic dressed photon research \cite{sak1}, Sakuma et al. \cite{sak3, sak4} shed new light on the long-forgotten GR theory (proved in axiomatic QFT) referred to in the introduction, which states that interactions among quantum fields must inevitably accompany spacelike momentum supports. Since these findings revealed a novel perspective on cosmology, especially for models of dark energy and dark matter, we will briefly review these findings before we discuss the abovementioned dark matter model $\tilde{\Omega}u^{\mu}$ further. First, on the basis of the GR theorem, in previous research, the mathematical form of a spacelike electromagnetic field, which can be regarded as the extension of the charge-free Maxwell equation into spacelike momentum domains, was investigated. The main conclusions can be summarized as follows ({\bf [i] - [iv]}): \\
\hrule width 127mm 
\vspace{5mm}
{\bf [i]} The extended electromagnetic 4-vector potential $U_{\mu}$ can be represented by the Clebsch parameterization (CP) \cite{shl} with the parameters ($\lambda, \phi$); the former satisfies the spacelike Klein-Gordon (KG) equation $\nabla_{\sigma}\nabla^{\sigma}\lambda - (\kappa_{0})^{2}\lambda =0$, where $\kappa_{0}$ is the experimentally determined dressed photon constant, while the latter satisfies either the same KG equation or $\nabla_{\sigma}\nabla^{\sigma}\phi=0$, depending on whether $U_{\mu}$ is spacelike or lightlike. The lightlike $U_{\mu}$ can be interpreted as a $U(1)$ gauge boson, while the spacelike $U_{\mu}$ provides the necessary spacelike momentum supports for field interactions. In fluid mechanics, CP is used for canonical $H$ formulations of barotropic fluids. CP is suitable for extended free Maxwell fields because, in sharp contrast to Eq. (\ref{eqn:33}), for the baroclinic case, the pseudoscalar $\Omega_{(ro)}$ defined by Eq. (\ref{eqn:67}) always vanishes:\begin{equation}
\Omega_{(ro)}:=S_{01}S_{23}+S_{02}S_{31}+S_{03}S_{12}=0,    \label{eqn:67}
\end{equation} 
where $S_{\mu\nu}:=\nabla_{\mu}U_{\nu}-\nabla_{\nu}U_{\mu}$ denotes the field strength of the extended electromagnetic field. According to Eq. (\ref{eqn:67}), as in the case of a free electromagnetic wave, the extended \lq\lq electric\rq\rq and \lq\lq magnetic\rq\rq fields are perpendicular to each other. \\

{\bf [ii]} The energy-momentum tensor $\hat{T}^{\mu\nu}$ for both cases can be written in a unified form as 
\begin{equation}
\hat{T}^{\mu\nu} = \hat{S}^{\mu\;\;\nu\sigma}_{\;\;\sigma}-\frac{1}{2}\hat{S}^{\alpha\beta}_{\;\;\;\;\alpha\beta}g^{\mu\nu},\;\;\;\hat{S}_{\alpha\beta\gamma\delta}:=S_{\alpha\beta}S_{\gamma\delta}.   \label{eqn:69}
\end{equation}
Note that due to the skew-symmetric nature of $S_{\mu\nu}$, $\hat{S}_{\alpha\beta\gamma\delta}$ satisfies exactly the same properties as the Riemann curvature tensor $R_{\alpha\beta\gamma\delta}$; that is,
\begin{eqnarray}
R_{\beta\alpha\gamma\delta}=-R_{\alpha\beta\gamma\delta},\;\;R_{\alpha\beta\delta\gamma}=-R_{\alpha\beta\gamma\delta},\;\;R_{\gamma\delta\alpha\beta}=R_{\alpha\beta\gamma\delta},   \\
R_{\alpha\beta\gamma\delta}+R_{\alpha\gamma\delta\beta}+R_{\alpha\delta\beta\gamma}=0.  \label{eqn:71}
\end{eqnarray} 
Eq. (\ref{eqn:71}) is known as the first Bianchi identity and corresponds to Eq. (\ref{eqn:67}). Therefore, $\hat{T}^{\mu\nu}$ in Eq. (\ref{eqn:69}) becomes isomorphic to the Einstein tensor $G^{\mu\nu}$, and its divergence vanishes. {\it Specifically, we can say that $\hat{T}^{\mu\nu}$ naturally fits into the geometrodynamics of general relativity.} \\

{\bf [iii]} Since the non-lightlike $U^{\mu}$ has a spacelike momentum field parameterized by $\kappa_{0}$ in the aforementioned spacelike KG equation, it forms a submanifold of de Sitter space (a pseudo-hypersphere $\mathfrak{D}$ embedded in $R^{5}$) with a geometrical structure similar to the spacelike KG equation, with the radius of $\mathfrak{D}$ corresponding to $\kappa_{0}$ as a scale parameter. According to Sakuma et al. \cite{sak3, sak4}, the importance of de Sitter space is twofold. First, using a spacelike momentum field in de Sitter space, Snyder \cite{hss} derived a spacetime quantization with a built-in Lorentz invariance; second, de Sitter space is a solution of the Einstein field equation, which describes the accelerated expansion of the universe. In accordance with Snyder\rq s work, they showed that the quantized form of $\hat{T}^{\mu\nu}$ can be given by a combined form of the Majorana fermion field, which behaves as an energy-momentum tensor with \lq\lq virtual photons\rq\rq acting as mediators of electromagnetic field interactions and $\hat{T}^{\mu\nu}$ associated with a unique ground state $M_{g}$, which can be regarded as a compound Rarita-Schwinger state with a spin of $3/2$. In terms of the accelerated expansion of the universe, they considered the possibility that the trace of the energy-momentum tensor representing $M_{g}$ can be interpreted as a \lq\lq reduced cosmological constant\rq\rq $\Lambda_{DP}$ (negative in our sign convention) whose magnitude can be evaluated by the new theory of dressed photons. In fact, $\Lambda_{DP}$ is $-2.47\times 10^{-53} m^{-2}$, while the observed value of $\Lambda_{obs}$, derived from Planck satellite observations \cite{hl}, is $-3.7\times 10^{-53} m^{-2}$.  \\

{\bf [iv]} The light field, including the newly identified spacelike counterpart $U_{\mu}$, is an essential element in the theory of dressed photons. An especially intriguing aspect of dressed photons is that even in the lightlike field $U_{\mu}$ that satisfies $(U_{\nu})^{*}U^{\nu}=0$, where $(U_{\nu})^{*}$ is the complex conjugate of $U_{\nu}$, the spacelike KG equation $\nabla_{\sigma}\nabla^{\sigma}\lambda - (\kappa_{0})^{2}\lambda =0$ is \lq\lq encoded\rq\rq in this lightlike field through CP. If the conformal symmetry of the lightlike $U_{\mu}$ field, which has $W^{2}=0$ and may be related to the Weyl curvature hypothesis proposed by Penrose \cite{rp2}, breaks, then the spacelike $U_{\mu}$ field emerges, along with the abovementioned $M_{g}$, which is responsible for generating de Sitter space. 
de Sitter space has the unique structural characteristic of twin universes \cite{sak4}, with each twin universe separated by the hypersurface of the event horizon embedded in it. Thus, starting from \lq\lq the big bang\rq\rq in the respective domains caused by the conformal symmetry breaking of the lightlike $U_{\mu}$ field, the twin universes (with one consisting of ordinary matter and the other consisting of anti-matter in the sense of time reversal) merge eons later at the event horizon and return to the original light phase. 
In this scenario, the creation and annihilation of the universe can be compared to those of elementary particles according to the intermediation of the light field. We believe that this cyclic twin universe cosmology is similar to the conformal cyclic cosmology \cite{rp3} proposed by Penrose. \\
\hrule width 127mm 
\vspace{5mm}
The newly proposed cosmology described above includes two noteworthy features that are missing from current cosmology based on cosmic inflation scenarios. First, in the former, instead of treating vacuum energy as starting from nothing, we assume the infinite cyclicity of the twin universes, with the \lq\lq nodes\rq\rq represented as a \lq\lq lightlike universe\rq\rq with null distance. One cycle begins with the conformal symmetry breaking of the nodal lightlike universe, and because of the property $W^{2}=0$\cite{sak4} (the aforementioned Weyl curvature hypothesis) of the nodal universe, the isotropy of the emerging non-lightlike universe can be naturally explained, in sharp contrast to inflation scenarios. Furthermore, the twin structure of the universe provides a simple solution to the missing anti-matter problem.

\subsubsection{On the flatness of the universe and the thermodynamic twin structure}
In addition to the isotropy of the universe, we must consider the flatness problem. Sakuma et al.\cite{sak3} proposed an SCSB hypothesis, which is referred to in the argument immediately before Eq. (\ref{eqn:63}). According to this hypothesis, the twin universes are assumed to be metric spacetimes emerging from simultaneous transitions from null electromagnetic and gravitational fields that satisfy $(U_{\nu})^{*}U^{\nu}=0$ and $W^{2}=0$, respectively, to the symmetry breaking spacelike $(U_{\nu})^{*}U^{\nu}<0$ and $W^{2}\neq 0$ fields. The key assumption of SCSB is that this transition can be parameterized by $\kappa_{0}$, and our reasoning is based on the following two facts: \\
(a) The abundance ratio of dark matter to dark energy is approximately $1/3$;\\
(b) The physical meaning of the cosmological term $\Lambda g^{\mu\nu}$ can be explained by Eq. (\ref{eqn:57}). \\
In subsection 5.2.1, we discussed a new plausible form of dark matter; however, given its overwhelming presence over ordinary matter, we do not believe that the dark matter modeled by $\tilde{\Omega}u^{\mu}$ is the most dominant form of dark matter. 

As we explained in item {\bf[iii]} of subsection 5.2.2, in the model of Sakuma et al., the observable effect of dark energy is generated by the reduced cosmological constant $\Lambda_{DP}$, which has been shown to be proportional to $3(\kappa_{0})^{2}$. We include the factor 3 in the expression of $\Lambda_{DP}$ because the spatial dimension of our universe is three. Recall that our original goal, which led to the introduction of $\Lambda_{DP}$, was to properly evaluate the involvement of the spacelike momentum field in quantum field interactions, and the existence of $\Lambda_{DP}$ was derived from electromagnetic field interactions as an important parameter characterizing the ground state of spacelike \lq\lq virtual photons\rq\rq. 
The similarity between Newtonian gravity and electromagnetic Coulomb force strongly suggests a possiblity that the physical entity similar to spacelike \lq\lq virtual photons\rq\rq must be involved in gravitational field interactions. Thus, we believe that it natural to assume that a similar argument can be extended to gravitational field interactions, although we do not have a satisfactory quantum gravitational theory. In electromagnetism, the spacelike extension of Maxwell theory is given by Eq. (\ref{eqn:69}). For the gravitational case, based on Eq. (\ref{eqn:65}), we introduce a spacelike extension with the form $\tilde{T}^{\mu\nu}=\tilde{\Omega}u^{\mu}u^{\nu}$, where $u^{\mu}$ satisfies the spacelike 4-vector condition $u^{\nu}u_{\nu}=-1$. We investigated this form because the important scalar parameter in our dark matter model must be identified in terms of the \lq\lq zero-point energy\rq\rq of the $\tilde{\Omega}$ field, which exists implicitly in the quantum version of Eq. (\ref{eqn:19}).

The energy-quantized version of $\tilde{T}^{\mu\nu}$ is the one in which $\tilde{\Omega}$ is discretized and the minimum value of $\tilde{\Omega}_{0}:=Min[|\tilde{\Omega}|]>0$ exists, which can be compared to the zero-point energy $h\nu/2$ of a harmonic oscillator. As previously stated in the discussion of dark energy by Sakuma et al.\cite{sak3}, the spacelike energy-momentum tensor $\tilde{T}^{\mu}_{\;\;\nu}=\tilde{\Omega}u^{\mu}u_{\nu}$ is not observable, except for its trace, which can be observed as invariant under general coordinate transformations. On the basis of their arguments, we may say that the zero-point energy $-\tilde{\Omega}_{0}$, as the trace of $(\tilde{T}^{\nu}_{\;\;\nu})_{0}$, can be reinterpreted by transferring it from the right-hand side of the Einstein field Eq. (\ref{eqn:61}) to the left-hand side as the {\it hypothetical cosmological term} $\Lambda(\tilde{\Omega}_{0}) g^{\mu\nu}$, in which the existence of $g^{\mu\nu}$ is formal and has no physical meanings, where\begin{equation}
4\Lambda(\tilde{\Omega}_{0}):=\frac{8\pi G}{c^{4}}\tilde{\Omega}_{0}.  \label{eqn:72}
\end{equation}
In contrast to the case of dark energy, however, we can investigate a different possibility in the dark matter model; that is, based on guiding fact (b), we can interpret $\Lambda(\tilde{\Omega}_{0})g^{\mu\nu}$ not as {\it the hypothetical cosmological term} mentioned above 
but instead as {\it \lq\lq a real\rq\rq cosmological term} with the form $\Lambda_{dm} g^{\mu\nu}$, where $\Lambda_{dm}:=\Lambda(\tilde{\Omega}_{0})$ and $g^{\mu\nu}$ represent the energy-momentum tensor of the conformal gravity field given by Eq. (\ref{eqn:57}). Then, by applying guiding fact (a) to $\Lambda_{dm}=\Lambda(\tilde{\Omega}_{0})$, we can introduce a core assumption of the SCSB hypothesis, which has the form 
\begin{equation}
\Lambda_{dm}= \Lambda(\tilde{\Omega}_{0}) = -\Lambda_{DP}/3>0.  \label{eqn:73}
\end{equation}  

First, we note that the condition of $W^{2}\neq 0$ in Eq. (\ref{eqn:57}) is necessary to define $g^{\mu\nu}$ as the energy-momentum tensor of the conformal gravity field, which allows us to hypothesize that $\Lambda_{dm}$ is directly related to the minimum value of quantized $|W^{2}|$. Recall that the magnitude of $\tilde{\Omega}_{0}$ depends on the undetermined constant parameter $T_{R}$, which was introduced in the argument of Eq. (\ref{eqn:63}). By adjusting this parameter, we can ensure that $\Lambda(\tilde{\Omega}_{0})$ is equal to $\Lambda_{dm}$, which corresponds to the minimum value of quantized $|W^{2}|$. Second, as previously mentioned, $-\Lambda_{DP}$ is proportional to $3(\kappa_{0})^{2}$; thus, according to Eq. (\ref{eqn:73}), the parameter $\Lambda_{dm}$ of the gravitational field is related to the parameter $(\kappa_{0})^{2}$ of the electromagnetic field.
Since the factor of 3 in $-\Lambda_{DP}$ reflects the spatial dimension of our universe, we hypothesize that $\Lambda_{dm} \propto (\kappa_{0})^{2}$ implies the \lq\lq equipartition of energy\rq\rq in four-dimensional spacetime, which can be concisely represented by a sign convention with the form $(+, -, -, -)$. The overwhelmingly large abundance ratios of dark energy and dark matter to ordinary matter suggest that the spacetime structure of our universe is determined by these two dark components.
As a result of the abovementioned \lq\lq equipartition of energy\rq\rq, our universe has a nearly flat spacetime structure. Thus, Eq. (\ref{eqn:73}) relates the parameter of conformal symmetry breaking in an electromagnetic field to that in a gravitational field in a way that it is consistent with observational evidence.

$\Lambda_{dm}$, as defined in Eq. (\ref{eqn:73}), is a unique parameter of anti-de Sitter space (ADS), the theoretical importance of which can be appreciated by ADS/CFT correspondence. Although an expansion-accelerated universe, such as ours, is not ADS itself, it is worth noting that the scale of our universe as one of the pair is given exactly by $\Lambda_{dm}$ \cite{sak4}; thus, in this sense, ADS must be a genuine scale parameter of our universe. 
If we choose a natural unit system in which the speed of light $c$ and the Planck constant $h$ have unit magnitude, then the length scale $\sqrt{1/(\kappa_{0})^{2}} \approx 40$ nanometer, called the dressed photon constant, can be shown to be the geometric mean \cite{sak4} of the smallest Planck length and the largest $\sqrt{1/\Lambda_{dm}}$. It is also worth noting that $\sqrt{1/(\kappa_{0})^{2}}$ provides a rough estimate of the Heisenberg cut for electromagnetic phenomena. We can argue that the dark matter phenomenon can be explained by combining $\tilde{\Omega}u^{\mu}$ currents in regions where $R^{\mu\nu}$ is negligibly small, which was noted in the argument after Eq. (\ref{eqn:65}), and the result derived of Eq. (\ref{eqn:73}).

Regarding the twin structure of universes and the thermodynamic aspect of spacetime, we briefly refer to Tomita-Takesaki theory \cite{tom} on the Kubo-Martin-Schwinger (KMS) state, which can be regarded as a generalization of the Gibbs state that covers thermodynamic equilibrium states with infinite degrees of freedom for which we cannot define trace operations. Since the KMS state is a mixed state, its corresponding Gel\rq fand-Naimark-Segal representation is reducible. Therefore, for $\mathcal{M}$ defined as a von-Neumann algebra in Hilbert space $\mathfrak{H}$, there exists a commutant $\mathcal{M}^{\prime}$ that satisfies the following inversion relation:
\begin{eqnarray}
J \mathcal{M}J = \mathcal{M}^{\prime},\;\;\; e^{itH}\mathcal{M} e^{-itH}=\mathcal{M},\;\;\;J^{2}=1,  \label{eqn:75} \\
J H J = -H,    \label{eqn:77}
\end{eqnarray}
where $H$ and $J$ denote the Hamiltonian and the anti-unitary operator known as the modular conjugate operator.
The spectrum of the Hamiltonian is symmetric with regard to its sign, indicating the existence of states with negative energy, whose stability is known to be greater than that of a vacuum state \cite{bw1}. Thus, we believe that the result of Tomita-Takesaki theory applies to the case of twin universes discussed in this article.

\section*{Acknowledgments}

This research was partially supported by a collaboration with the Institute of Mathematics for Industry, Kyushu University.

\appendix

\section{}
Dividing Eq. (\ref{eqn:27}) by $n$, we have
\begin{equation}
(w/n)u^{\nu}\partial_{\nu}u_{\mu}-(\partial_{\mu}p)/n+(u_{\mu}u^{\nu}/n)\partial_{\nu}p =0.   \label{app1}
\end{equation}
The first term on the left-hand side of (\ref{app1}) can be rewritten as
\begin{eqnarray}
u^{\nu}(w/n)\partial_{\nu}u_{\mu} &=& u^{\nu}[\partial_{\nu}\{(w/n)u_{\mu}\}-u_{\mu}\partial_{\nu}(w/n)] \nonumber \\
&=& u^{\nu}[\partial_{\nu}\{(w/n)u_{\mu}\}-\partial_{\mu}\{(w/n)u_{\nu}\}+\partial_{\mu}\{(w/n)u_{\nu}\}] \nonumber \\
 &-& u^{\nu}u_{\mu}\partial_{\nu}(w/n).   \label{app2}
\end{eqnarray}
With the new notation for the vorticity field $\omega_{\mu\nu}:=\partial_{\mu}[(w/n)u_{\nu}]-\partial_{\nu}[(w/n)u_{\mu}]$, (\ref{app2}) becomes
\begin{eqnarray}
u^{\nu}(w/n)\partial_{\nu}u_{\mu} = -\omega_{\mu\nu}u^{\nu}+u^{\nu}\partial_{\mu}\{(w/n)u_{\nu}\}
 - u^{\nu}u_{\mu}\partial_{\nu}(w/n).   \label{app3}
\end{eqnarray}
By substituting (\ref{app3}) into (\ref{app1}), we obtain 
\begin{equation}
-\omega_{\mu\nu}u^{\nu}+u^{\nu}\partial_{\mu}\{(w/n)u_{\nu}\}-(\partial_{\mu}p)/n-u_{\mu}u^{\nu}[\partial_{\nu}(w/n)-(\partial_{\nu}p)/n] =0.  \label{app4}
\end{equation}
Next, by applying $u^{\nu}u_{\nu}=1$ to the second term in (\ref{app4}), we obtain
\begin{equation}
u^{\nu}\partial_{\mu}\{(w/n)u_{\nu}\}=\partial_{\mu}\{(w/n)u^{\nu}u_{\nu}\}-(w/n)u_{\nu}\partial_{\mu}u^{\nu}=\partial_{\mu}(w/n).   \label{app5}
\end{equation}
In addition, according to Eqs. (\ref{eqn:17}) and (\ref{eqn:23}), the sum of the last two terms in (\ref{app4}) becomes $-u_{\mu}u^{\nu}T\partial_{\nu}(\sigma/n)=0$; thus, we can finally obtain that 
\begin{equation}
-\omega_{\mu\nu}u^{\nu}+\partial_{\mu}(w/n)- (\partial_{\mu}p)/n =0,\;\;\Longrightarrow\;\; 
\omega_{\mu\nu}u^{\nu}=T\partial_{\mu}(\sigma/n).   \label{app6}
\end{equation}

\end{document}